\documentclass[preprint,journal]{vgtc}

\ifpdf
  \pdfoutput=1\relax                   
  \usepackage{graphicx}                
  \DeclareGraphicsExtensions{.pdf,.png,.jpg,.jpeg} 
\else
  \usepackage{graphicx}                
  \DeclareGraphicsExtensions{.eps}     
\fi%

\graphicspath{{figs/}{figures/}{pictures/}{images/}{./}} 

\usepackage{microtype}                 
\PassOptionsToPackage{warn}{textcomp}  
\usepackage{textcomp}                  
\usepackage{mathptmx}                  
\usepackage{times}                     
\usepackage{cite}                      
\usepackage{tabu}                      
\usepackage{booktabs}                  

\usepackage{mathptmx}                  

\usepackage{csquotes}
\usepackage{amsmath}
\usepackage{amssymb}
\usepackage{metalogo}
\usepackage{nicefrac}
\usepackage[inline]{enumitem}

\usepackage[svgnames]{xcolor}

\usepackage{customlabel}

\RequirePackage{hyperref}
\usepackage[hyphenbreaks]{breakurl}
\usepackage[capitalize]{cleveref}

\onlineid{0}

\vgtccategory{Research}

\vgtcinsertpkg

\vgtccategory{Research}

\vgtcpapertype{algorithm/technique}

\title{Compact Phase Histograms for Guided Exploration of Periodicity}

\author{%
  Max Franke\thanks{Both authors are with the University of Stuttgart, Germany.\phantom{filler  break}
    \indent\phantom{\;*}E-mail: \texttt{\{Max.Franke,Steffen.Koch\}@vis.uni-stuttgart.de}}
  \and
  Steffen Koch*
}

\authorfooter{
  \item Max Franke and Steffen Koch are with the University of Stuttgart, Germany.
    E-mail: \{First\}.\{Last\}@vis.uni-stuttgart.de
  }

\abstract{%
Periodically occurring accumulations of events or measured values are present in many time-dependent datasets and can be of interest for analyses.
The frequency of such periodic behavior is often not known in advance, making it difficult to detect and tedious to explore.
Automated analysis methods exist, but can be too costly for smooth, interactive analysis.
We propose a compact visual representation that reveals periodicity by showing a phase histogram for a given period length that can be used standalone or in combination with other linked visualizations.
Our approach supports guided, interactive analyses by suggesting other period lengths to explore, which are ranked based on two quality measures.
We further describe how the phase can be mapped to visual representations in other views to reveal periodicity there.
}

\ieeedoi{10.1109/VIS54172.2023.00047}

\teaser{
\centering
\includegraphics[scale=1]{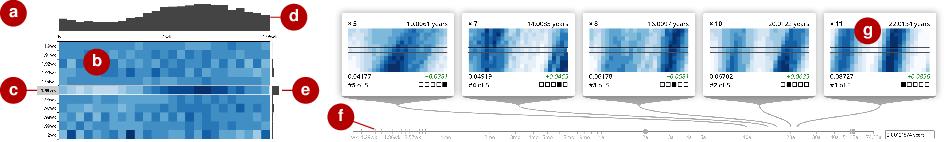}
\caption{%
  Our approach is centered around the phase histogram widget~\textbf{(a)}, which indicates the von Mises distribution~\cite{vonMises_1918} for given period lengths as a row-wise heat map~\textbf{(b)}.
  The duration of the period can be adjusted by vertically scrolling the heat~map.
  The current period length is indicated by a frame~\textbf{(c)} and visualized additionally as a bar chart~\textbf{(d)}.
  The quality measures for period lengths are visualized as a vertical bar chart~\textbf{(e)}.
  Bigger changes of the period length or picking a dedicated one is possible with the time slider~\textbf{(f)}.
  Other interesting period lengths that are fractional multiples of the current one are
  suggested above~\textbf{(g)} the slider.
}
\label{fig:teaser}
\customlabel{fig:teaser:widget}{a}
\customlabel{fig:teaser:heatmap}{b}
\customlabel{fig:teaser:current-row}{c}
\customlabel{fig:teaser:histogram}{d}
\customlabel{fig:teaser:quality-histogram}{e}
\customlabel{fig:teaser:slider}{f}
\customlabel{fig:teaser:suggestions}{g}
}

\CCScatlist{
  \CCScatTwelve{Human-centered computing}{Visu\-al\-iza\-tion}{Visu\-al\-iza\-tion techniques}{}
}

\begin{document}
\firstsection{Introduction}
\maketitle
\label{sec:introduction}

Many datasets contain temporal information.
For many years, visualization techniques have been used to understand developments, trends, and other temporal patterns in such time data series~\cite{Aigner_2011}.
Such patterns include periodically occurring accumulations of events, measured values, or frequencies.
In time series data, one or more attributes depend on time: $a(t)$.
Accordingly, periodic behavior of period length $\tau$ can be seen as a similarity of the characteristics of $a(t)$ in the intervals $\left[ t_0 + k \tau,~ t_0 + k \tau + \tau \right],~ k \in \mathbb{Z}$.
A special case are event data, where the temporal distribution of data points is of interest, rather than the distribution of values over time.
Periodic behavior can be interesting for analyses;
for instance, to predict future values of a data attribute, or to find hidden dependencies in the data.
Several well-known automated procedures for detecting periodic effects exist, such as Fourier transforms~\cite{Brigham_1967}, seasonal-trend decomposition based on Loess~(STL)~\cite{Cleveland_1990}, and dynamic mode decomposition~(DMD)~\cite{Schmid_2010,Krake_2022}.
However, some of these methods require the period length of interest as an input parameter, or can be expensive regarding the compute time.

In certain situations, the \emph{interactive} visual identification and exploration of periodic effects has benefits over automatic procedures and static visual approaches;
for example, if the interplay of periodicity in events with their geographical location or other data attributes is of interest.
Different periodicities can be explored without making presumptions on interesting periods.
In addition, important facets of the time-dependent data become visible immediately.
Various approaches~\cite{Weber_2001,vanWijk_1999,Brehmer_2017,Kosara_2011,Sips_2012} map phase and time to positions in a grid- or spiral-based layout, revealing periodicity in an intuitive manner.
We extend previous work with a compact, aggregated, and interactive representation (\cref{fig:teaser:widget}) of time series data that reveals periodic behavior even for datasets with larger temporal extent.
We offer guidance~\cite{Ceneda_2019} in the form of suggestions (\cref{fig:teaser:suggestions}) and quality measures~\cite{Shannon_1948,vonMises_1918}.
Since exploring the interplay with other data attributes can be interesting, our approach can be integrated within a larger, multiple-view visualization.

The contributions of the approach presented in this work comprise the introduction of a new composite widget that
\begin{enumerate*}[label=(\roman*)]
  \item helps users detect and explore periodic occurrences of aggregated quantitative information in long time series data,
  \item offers guidance to point users to potentially interesting periods exhibiting seasonal effects, and
  \item lets users pick suitable mappings including glyphs and color scales to understand periodicity in other views.
\end{enumerate*}

\section{Related Work}
\label{sec:related-work}

We briefly discuss related automatic methods for analyzing periodic behavior, but focus on interactive visual approaches.
Various possibilities for representing cyclical temporal data exist~\cite{Aigner_2011,Brehmer_2017}.
Hence, we limit our discussion here to approaches closely related to ours.

\paragraph*{Automatic Periodicity Analysis.}
Fourier analysis~\cite{Brigham_1967} converts an input signal to the frequency domain.
This requires a high sampling rate on the input signal, which can be expensive for event data, where it must be generated as a long, sparse histogram of the events.
Fourier analysis also returns many false positives for non-sinusoidal periodicity.
Cleveland et al.~\cite{Cleveland_1990} introduced STL, which splits a time-dependent data signal into a seasonally recurring component, a linear trend, and a remainder.
Constrained DMD~(cDMD)~\cite{Krake_2022} produces similar results using a different method.
Cycle plots~\cite{Cleveland_1993,Bogl_2017} produce visualizations that also reveal such seasonal characteristics.
While these methods produce good results even for noisy data, they are usually expensive to compute, and require prior knowledge on the frequency of the periodic behavior.
STL also requires the period length to be an integer multiple of the sampling rate.
Unconstrained DMD~\cite{Schmid_2010} does not require the period length as input, but has similar computation costs as cDMD~\cite{Krake_2022}, and does not always output the expected period length.
Fourier analysis~\cite{Brigham_1967}, DMD~\cite{Schmid_2010}, and cDMD~\cite{Krake_2022} find sinusoidal signal components; but struggle with other signal characteristics that could occur with event data.
In contrast to these methods, we focus on interactive and exploratory analysis of periodic behavior in event data, often in the context of other data attributes.

\paragraph*{Visualization of Periodicity.}
Various works have explored non-linear layouts of data to reveal periodicity.
Carlis et al.~\cite{Carlis_1998} and Weber et al.~\cite{Weber_2001} proposed Archimedean spirals, where one turn along the spiral represents one period.
Periodic behavior of that period length would then appear as lines or cones going radially outward from the center (\cref{fig:patterns:spiral}).
The same effect can be reached with a line-wise, rectangular representation (\cref{fig:patterns:rect,fig:patterns:rect-multiple,fig:patterns:rect-fraction}), where each line represents one period~\cite{Kosara_2011}.
In both representations, spiral and rectangular, the period length can be adjusted interactively, by tightening the spiral or changing the aspect ratio of the rectangle, to find periodicities.
Our approach focuses particularly on scalability, as well as guidance of users to interesting period lengths.
We show these representations as detail views on demand.
A special case of the rectangular representations are calendar-based layouts~\cite{vanWijk_1999,Lammarsch_2009,Silva_2021}, which reveal periodic behavior in human-made time concepts on multiple scales, but are limited to fixed time concepts.
Several visualizations utilizing concentric circles~\cite{Argyriou_2012,Bale_2007,Hao_2013,Mariano_2018} or stacks~\cite{Lee_2010b} also fit this category.
Frey et al.~\cite{Frey_2012} present a matrix representation which reveals self-similar patterns in time series data, but requires considerable screen estate.

\emph{Pinus} view~\cite{Sips_2012} shows an aggregated visual summary of a time series in a triangular fashion, where each position in the triangle corresponds with an interval and an offset in the temporal domain of the dataset.
The residuality model by Van de Weghe et al.~\cite{Van_de_Weghe_2007} describes a similar concept.
These concepts, like ours, show a visual summary of the temporal domain.
However, they are orthogonal to ours in that they consider aggregated intervals, rather than looking at repeating patterns.
Closer to our approach is the work by Suschnigg et al.~\cite{Suschnigg_2021}, who compare characteristics between periods via glyphs, or a matrix of anomaly scores.
The approach by Ishii and Misue~\cite{Ishii_2018} is also closely related to our quality metrics, as they essentially visualize the von Mises~\cite{vonMises_1918} vector strength and direction for different data attributes.
Recurrence quantification analysis~\cite{Marwan_2007,Rawald_2017} also share similarities to our work.
In contrast to these works, we focus on interactive discovery of interesting period lengths, and offer a more compact visual representation.

\section{Approach}
\label{sec:approach}

Our approach is centered around an aggregated view on all periods at once for a given period length, which we call the \emph{phase histogram.}
Periodic behavior for that period length is visible by non-uniform distribution of values (Figs.~\ref{fig:teaser:heatmap}, \ref{fig:teaser:histogram}, \ref{fig:patterns:ours-sharp}, and \ref{fig:patterns:ours-fuzzy}; \cref{tab:example-measure-values}).
We calculate quality measures for different period lengths to guide towards interesting ones, focusing on fractional multiples of the current period length to balance out the challenges discussed in \cref{subsec:guidance}.

\subsection{Phase Histogram}
\label{subsec:phase-histogram}

Our approach considers a time series of events $S$, $\left| S \right| = n$.
The time series $S = \left\{ t_i ~\middle|~ i \in \left\{1, \dots, n\right\} \subset \mathbb{N} \right\}$ consists of points in time $t_i$.
It has an extent from $t_0$ to $t_1$: $t_i \in \left[ t_0, t_1 \right]$, where $t_0, t_1 \in \mathbb{R}$.
For a user-selectable \emph{period length} $\tau$, each point in time $t$ has a \emph{phase} $\varphi(t, \tau) \in \left[0, 2\pi\right)$.
The phase is the offset from the start of the period relative to the period length, where the first period starts at $t_0$:
\begin{center}
  $\varphi(t, \tau) = 2 \pi \cdot \frac{\left( t - t_0 \right) ~\mathrm{mod}~ \tau }{\tau}$
\end{center}

\begin{figure*}[tb]
  \begin{center}%
    \includegraphics[scale=1]{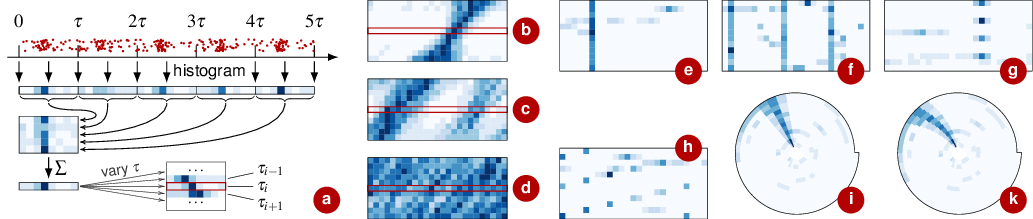}
  \end{center}%

  \caption{
    A schematic explanation~\textbf{(a)} of our approach:
    Event data gets binned over the temporal domain.
    For the Cartesian representations~\cite{Kosara_2011,vanWijk_1999,Lammarsch_2009,Silva_2021} the bins then get placed so that each row represents one period.
    Our approach shows an aggregated view on this, and varies the period durations slightly from row to row.
    The current period length is framed in red in the figure.
    Example patterns for periodic behavior in our approach for a sharp~\textbf{(b)} and a less sharp~\textbf{(c)} periodic pattern, and for uniform noise~\textbf{(d)}.
    As a comparison,
    periodic behavior~\textbf{(e)} in the Cartesian binned representation is shown for the actual signal period length, and for
    multiples~\textbf{(f)},
    integer fractions~\textbf{(g)}, and non-integer fractions (\textbf{h,} here $\nicefrac{6}{5}$) thereof.
    Here, the periodicity manifests as vertical lines.
    Patterns appear as straight lines emanating from the center in the Archimedean binned representation~\textbf{(i)}~\cite{Carlis_1998,Weber_2001}, but are also visible if the period length is nearly right~\textbf{(k)}.
  }
  \label{fig:patterns}
  \customlabel{fig:patterns:schematic}{a}
  \customlabel{fig:patterns:ours-sharp}{b}
  \customlabel{fig:patterns:ours-fuzzy}{c}
  \customlabel{fig:patterns:ours-random}{d}
  \customlabel{fig:patterns:rect}{e}
  \customlabel{fig:patterns:rect-multiple}{f}
  \customlabel{fig:patterns:rect-fraction}{g}
  \customlabel{fig:patterns:rect-real-fraction}{h}
  \customlabel{fig:patterns:spiral}{i}
  \customlabel{fig:patterns:spiral-skewed}{k}
\end{figure*}

For a given $\tau$, we then calculate a histogram with $N$ bins over the phases of all time series events (\cref{fig:patterns:schematic}).
The choice of $N$ depends mainly on available space.
It affects the granularity of the resulting phase histogram, as well as the scaling of the entropy estimation values.
In the subsequent examples, we use $N=25$.
The shape of the histogram then reveals potential periodic behavior if one or some of the bins contain considerably more items than the others.
Figures~\hyperref[fig:patterns]{2b, c, e to k}, and \cref{tab:example-measure-values} (rows 2--5) show some examples of histograms which represent periodic behavior.
Multiples and fractions of the actual periodic behavior's period length will also produce interesting-looking patterns.
We consider such multiples and fractions to evaluate whether they show clearer periodic behavior than the current period length $\tau$ (see~\cref{subsec:guidance}).
While our approach focuses on event data, it can also be employed to show periodic behavior of time-dependent data attributes (\cref{subsec:visual-representation}).
Our representation can be understood as a vertical aggregation of the rectangular binned representations~\cite{Kosara_2011,vanWijk_1999,Lammarsch_2009,Silva_2021}, which reveals overarching patterns in all rows (see~\cref{fig:patterns:schematic}).

\subsection{Pre-calculation and Guidance}
\label{subsec:guidance}

\begin{table}[tb]
  \caption{
    Example phase histograms alongside the two quality measures used:
    Shannon entropy~\cite{Shannon_1948} and von Mises vector strength~\cite{vonMises_1918}.
    For the vector strength, the distribution of the data points on the unit circle (blue) and their barycenter (red) are also visualized.
  }
  \label{tab:example-measure-values}
  \begin{center}
    \includegraphics[scale=1]{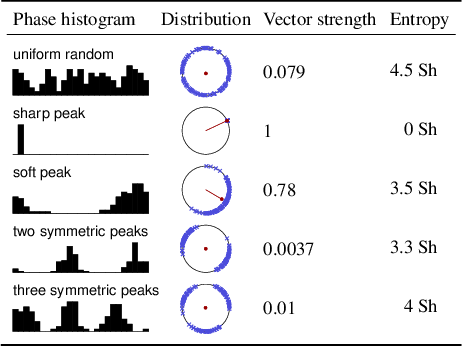}
  \end{center}
\end{table}

Interesting period lengths are those that match periodically reoccurring events in the data.
In these cases, the phases of a larger-than-average number of events are similar, and the phase histogram is not uniform.
Our approach calculates two quality measures that help decide which period lengths show promising patterns.
The first measure is the Shannon entropy~\cite{Shannon_1948} of a phase histogram.
Shannon entropy is high for more uniformly distributed histograms, and low for histograms with clear peak outliers~(\cref{tab:example-measure-values}).
The second measure is von Mises vector strength~\cite{vonMises_1918}, which is a measure between 0 and 1, and is highest when the phases of all data points are the same.
Von Mises vector strength is approximated by the distance from the center of a unit circle to the barycenter of all data points after these have been projected onto the unit circle at the angle of their respective phase.
Both measures are suited for the task, since they generally indicate when the phase histograms are not uniformly distributed.
They are also fast to compute, and so can be used as part of a highly interactive exploration approach.
They have individual downsides, which is why we calculate both for a more nuanced view on the data:
For Shannon entropy, the values quickly rise with only little noise.
Von Mises vector strength cannot recognize whole multiples of the period length of periodic behavior, as the barycenter then moves to the center of the unit circle (\cref{tab:example-measure-values}, last two rows).

For a dataset, our approach initially pre-calculates the histograms and quality measures for a set of period lengths between a lower bound (e.g., 1\,min) and an upper bound (the temporal extent of the dataset).
Within this range, multiples of time units are sampled ($1 \dots 59\,\mathrm{min}$, $1 \dots 23\,\mathrm{h}$, etc.).
In addition, exponentially increasing period lengths are included.
Data for additional period lengths can then be requested during analysis, and are calculated ad-hoc.

We also offer suggestions based on the current period length as a guidance.
A key challenge here is to determine the actual frequency or period length of periodic behavior:
A repeating pattern of period length $\tau$ looks repeating also for multiples of $\tau$ (\cref{tab:example-measure-values,fig:patterns:rect-multiple}), but also for fractions thereof, if only an aggregation of all periods is considered (\cref{fig:patterns:rect-fraction}).
Furthermore, fractions of multiples (such as $\nicefrac{4}{3^\mathrm{rds}}$ of the actual period length) can look interesting as well in an aggregated view on the data (\cref{fig:patterns:rect-real-fraction}).
To facilitate discovery of better period lengths that are located at such factors, our approach samples a small set of fractions $\nicefrac{k}{n}$, where $n$ is a small natural number $\ge 2$ and $k \in \left\{ 1, \dots, 2n - 1\right\}$.
Small natural-numbered multiples of the current period length are sampled as well, and a selection of the best-ranking suggestions are returned based on available space.
This sampling happens on the fly for the current period length $\tau$.

\subsection{Visual Representation}
\label{subsec:visual-representation}

We propose a compact widget (\cref{fig:teaser:widget}) that can be included as part of a larger, coordinated-views visualization.
The widget shows the aggregated phase distribution for the currently selected period~(\cref{fig:teaser:histogram}) length $\tau$, as well as for the immediate neighborhood period lengths.
This context (\cref{fig:teaser:heatmap}) is shown as a pixel-based~\cite{Keim_2000} visualization or heat map, where each row shows a period length.
Each row visualizes the histogram as a line of colored rectangles, where the bin's respective value is mapped to color.
The period lengths are shown in ascending order from the pre-calculated data (\cref{subsec:guidance,fig:patterns:schematic}), with a context of $n$ rows above and below the current row, which is located at the center of the heat map~(\cref{fig:teaser:current-row}).
By scrolling on the heat map with the mouse wheel, or clicking on rows, users can browse the rows, changing the current period length accordingly.
By default, the color of the heat map cells represents the number of data items in that histogram bin.
It is also possible to map this to other measures calculated on these subsets of data, such as the mean value or variance of another data attribute (shown in \cref{fig:teaser:suggestions} for the number of sun spots per day).

To the right of the heat map, an interest measure is indicated for each row by a vertical bar chart (\cref{fig:teaser:quality-histogram}).
For the vector strength (\cref{subsec:guidance}), which is shown by default, the measure is mapped directly.
For the entropy, the highest possible entropy (least interesting) is mapped to an empty bar, and the lowest possible entropy (most interesting) is mapped to a full bar.
At the top of the widget, the current period length's histogram is visualized again, as a bar chart (\cref{fig:teaser:histogram}).
In applications where the phase is mapped to visual attributes in other views (see~\cref{subsec:mapping-phase}), this visual mapping is displayed in a legend (\cref{fig:mapping:legend-color,fig:mapping:legend-glyph}) below the bar chart.

Our approach also contains a time slider~(\cref{fig:teaser:slider}), which represents the domain of possible period lengths on a logarithmic scale and can be used for larger adjustments to the current period length $\tau$.
Period lengths with the highest-ranked quality measure values are indicated as longer tick marks to guide users.
During exploration, multiples and fractions for the current period length are sampled.
The sampled period lengths are ranked by the selected quality measure, and the most promising ones are suggested as small thumbnails (\cref{fig:teaser:suggestions}) above the time slider.
Clicking on one such thumbnail changes the current period length to that of the thumbnail.

\subsection{Visually Mapping the Phase}
\label{subsec:mapping-phase}

\begin{figure}[t]
  \begin{center}
    \includegraphics[scale=1]{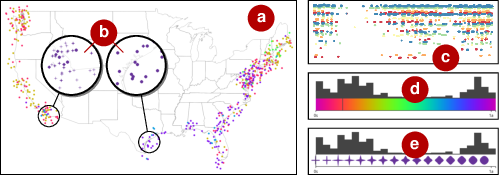}
  \end{center}

  \caption{
    Example mappings of phase to color~\textbf{(a,\,c)} or shape of visual marks~\textbf{(b)} in a scatter plot.
    Periodic behavior related to the spatial aspect of the data is revealed by uniform areas.
    The mapping is shown in our widget~(\cref{fig:teaser:widget}) as a legend~\textbf{(d,\,e)} that can be interactively adjusted to change the mapping of glyph or color to phase.
  }
  \label{fig:mapping}
  \customlabel{fig:mapping:color}{a}
  \customlabel{fig:mapping:shape}{b}
  \customlabel{fig:mapping:calendar-honolulu}{c}
  \customlabel{fig:mapping:legend-color}{d}
  \customlabel{fig:mapping:legend-glyph}{e}
\end{figure}

The presence of periodicity in the context of other data attributes can be of interest as well.
For example, users might be interested in understanding whether certain events displayed on a map occur at recurring times (see~\cref{sec:case-study}).
We have tested this mapping with position-based visualizations in our approach, but it could be extended to other visualization types.
Our prototype implementation contains a scatter plot (\cref{fig:mapping:color}), which represents the spatial attribute of the data.
To reveal spatio-temporal periodicity patterns, we propose two different types of mapping that can be applied to the markers of the scatter plot: color and shape (\cref{fig:mapping}).

The first variant maps the phase of the data item to a position on a continuous color scale.
Cyclical color scales, such as a rainbow color scale, can be a reasonable choice in this case.
However, some datasets might favor color scales with a clear cut between the end and the beginning.
The second variant maps to a visual mark with a parameterized shape.
For our prototype, we offer a mapping to a moon phase-like mapping, or the angular rotation within $\left[ 0, \pi \right)$ of a rectangle, both cyclical.
As acyclic mappings, we offer star shapes that morph into circles.
Previous works~\cite{Ebert_2000,Ropinski_2011,Borgo_2013} on glyph design offer additional options here.
Future work might explore suitable shapes for representing cyclical data attributes.

Within our central widget, we show a legend (\cref{fig:mapping:legend-color,fig:mapping:legend-glyph}) of this mapping.
Especially for the acyclic mappings, it can be beneficial to change the offset of the mapping.
Hence, the offset can be adjusted from within the legend.

\section{Case Study}
\label{sec:case-study}

We tested our approach with various synthetic and real-world datasets.
Here, we demonstrate its applicability on tidal sea level data obtained from NOAA~\cite{NOAA-Data}, with two specific datasets.
Such data contains various periodicities pertaining to the rotation of the earth in relation to the moon, the orbit of the moon around the earth and the earth around the sun, but also to annual or perennial patterns stemming from meteorological phenomena.
The interplay of these factors is not always obvious, demonstrating the need for exploratory analysis of the resulting patterns in the data.

The first dataset was generated from hourly measurements of the mean sea level (MSL) at one station located in Honolulu, Hawai'i~\cite{Honolulu-Data}.
The station measurements cover 118 years, from 1905 to the present day.
We reduced this data to the measurements that exceeded a threshold value ($+0.5\,\mathrm{m}$), leaving us with 5060 events.
Our approach guided us to find the spring and neap tide periodicity at half a sidereal month ($13.66\,\mathrm{d}$, \cref{fig:teaser:widget}).
We also could see and verify the $18.61$-year nodal cycle discussed by Haigh et~al.~\cite{Haigh_2011}.
By mapping time of year to the $x$ and year to the $y$ coordinate in a scatter plot~(\cref{fig:mapping:calendar-honolulu}, days increase to the right, years to the top), we can also see an increase of events over the years, and a seasonal yearly component (denser and less dense vertical areas).

The second dataset contains the dates with highest extreme water levels for all observation stations, as provided by NOAA~\cite{NOAA-Data}.
Over a record interval of 122.5 years, this data only contains 571 events, distributed over 74 unique stations.
Hence, the observed patterns are not as expressive;
additional preprocessing of raw data with more domain knowledge might yield better results here.
Still, we can see some periodicity at the $18.61$-year period length~\cite{Haigh_2011}.
Furthermore, seasonal differences in events between the US east and west coast are clearly visible in the mapping of phase to color in a geographical scatter plot~(\cref{fig:mapping:color}) for a one-year period length.

\section{Discussion}
\label{sec:discussion}

We implemented a web-based prototype for our approach.
Its backend calculates the required data, as well as additional detail data on demand.
We discuss initial results, and benefits and disadvantages compared to other visual representations and automated methods.

\paragraph*{Suitable Datasets and Use Cases.}
Our approach can be applied to use cases where the periodic re-occurrence of events is of interest.
Such events can also be generated from general time series data by determining points in time where time-dependent data values match some criteria.
We demonstrated the viability of this strategy for finding periodic behavior in real-world data in our case study~(\cref{sec:case-study}).
We offer a few sample datasets in our prototype, but also allow for arbitrary datasets to be loaded by the user.

\paragraph*{Implementation Details and Scalability.}
Our web-based prototype visualizes pre-calculated data, and additional data is calculated on demand asynchronously in a backend.
With this strategy, even for larger datasets with tens of thousands of events, interaction with the user interface is very responsive and smooth, with no or very few frame drops.
Suggestions are calculated ad-hoc by the backend and typically appear within half a second after the last interaction.
The thumbnails at the top (\cref{fig:teaser:suggestions}) work well to guide users to interesting period lengths.
We have found that the two measures supplement each others' drawbacks (\cref{subsec:guidance}) quite well.
The mapping of phase to color or shape in the scatter plot has proven useful to detect periodic behavior that is local in other data attributes (\cref{fig:mapping}).
For performance, redrawing the scatter plot is the bottleneck, but we have found datasets with thousands or tens of thousands of data points to still render interactively.
The implementation is available on GitHub and Zenodo~\cite{github}.

\paragraph*{Comparison to automated analyses.}
STL~\cite{Cleveland_1990} and related methods~\cite{Krake_2022} allow for the automated analysis of periodic behavior, and handle noisy data better.
These methods require prior knowledge of the period length for which the periodic behavior should be analyzed.
We compared mean computation time over ten runs of our method, STL, and DMD for two datasets:
For a smaller synthetic dataset, STL took $227.2\,\mathrm{ms}$ to calculate for \emph{one} period length of interest, and $3.45\,\mathrm{s}$ for the Hawai'i tide dataset~\cite{Honolulu-Data}.
DMD took $2.53\,\mathrm{s}$ and $1.63\,\mathrm{s}$, respectively, to produce decompositions with six components and a sensible delay parameter, but did not find the expected periodic behavior.
In contrast, our method took $470\,\mathrm{ms}$ and $589.5\,\mathrm{ms}$, respectively, to calculate the phase histograms and quality measures for over $1800$ sampled period lengths.
  The fairly high computation costs of methods like STL or cDMD prohibit extensive pre-calculations, or on-the-fly computation during smooth interaction.
  We also see that DMD struggles to produce good results for non-sinusoidal periodic behavior.
However, these methods could still be utilized as a second step to verify results found interactively with our approach on demand.
Fourier analysis~\cite{Brigham_1967} is another option, but not always suitable to real-world periodicity patterns.
We argue that automated analysis is most useful for targeted use cases, but that exploratory analysis, especially in the context of other data attributes (\cref{subsec:mapping-phase}), still requires more interactive methods.

\paragraph*{Comparison to other visual representations.}
Cycle plots~\cite{Cleveland_1993,Bogl_2017} are a powerful way to visualize periodicity, but again presume knowledge of the periodic behavior.
The calendar-based representations~\cite{vanWijk_1999,Lammarsch_2009} presume specific period lengths as well, but are often suitable for data based on from human-specified time concepts, such as months.
We think other rectangle-\cite{Kosara_2011} and spiral-based~\cite{Carlis_1998,Weber_2001} visualizations show periodic behavior in an intuitive manner, especially when period length can be adjusted.
However; these representations do not scale well for larger time spans; and interactive use can often lead to large shifts, bad aspect ratios, and flickering effects.
Our approach condenses these two-dimensional representations down to a one-dimensional aggregation.
This opens up space for visualizing different data attributes to understand relations between them, and allows showing close-by period lengths as a context to the current one.
This supports users in finding the best local period for periodic events.
The aggregation hides periodicity that only appears for a part of the temporal extent, and multiples or fractions of the actual periodicity are not clearly identifiable as such (\cref{subsec:guidance}).
Hence, we offer guidance, as well as detailed views as a tooltip to closer examine interesting period lengths (\cref{fig:patterns}).

\paragraph*{Open challenges.}
We demonstrated the general applicability of our approach with a prototype and different datasets, and plan to explore its integration into larger visualization systems.
Understanding the best choices with respect to phase mapping (\cref{subsec:mapping-phase}) in different scenarios, and determining the most appropriate visual mapping in terms of color scales and glyph designs to represent periodicity, is another interesting research direction we plan to pursue.
The integration of other quality measures such as variance of the phase, domain-specific measures, or the combination of multiple measures, could be an interesting extension.
Incorporating more complex automatic analysis methods, such as STL, could be another good addition to verify promising findings on demand.

\section{Conclusion}
\label{sec:conclusion}

We have presented a novel, aggregated visual representation of temporal data to explore periodicity.
The time domain is mapped to the phase or remainder for a given and adjustable period length and visualized in a binned manner.
Patterns that repeat with the same period length appear clearly in our representation.
Our approach offers a more compact, interactive alternative to existing representations.
It can be employed as a part of visualization systems with many views to let users configure the mapping of phases for depicting periodicity in other visual contexts.

\section*{Acknowledgments}

This work has been partially funded by the German Research Foundation (DFG) project \#\,314647693.
We thank Tim Krake in particular; as well as Frank Heyen, Daniel Klötzl, and Kuno Kurzhals; for their valuable feedback and assistance.

\bibliographystyle{abbrv-doi-hyperref}
\bibliography{bibliography}
\end{document}